\documentclass[journal=ancac3]{achemso}
\usepackage[version=3]{mhchem}
\usepackage{bm}

\author{Devynn M. Wulstein}
\affiliation{Department of Physics and Biophysics, University of San Diego, San Diego, CA 92110, USA}
\author{Ryan McGorty}
\affiliation{Department of Physics and Biophysics, University of San Diego, San Diego, CA 92110, USA}
\email{rmcgorty@sandiego.edu}

\title{Point-spread function engineering enhances digital Fourier microscopy}

\begin{document}
%% To be edited by editor
% \dates{Compiled \today}

%\ociscodes{(180.0180) Microscopy; (110.7348) Wavefront encoding; (070.0070) Fourier optics and signal processing.}

%% To be edited by editor
% \doi{\url{http://dx.doi.org/10.1364/XX.XX.XXXXXX}}

\begin{abstract}
While numerous optical methods exist to probe the dynamics of biological or complex fluid samples, in recent years digital Fourier microscopy techniques, like differential dynamic microscopy, have emerged as ways to efficiently combine elements of imaging and scattering methods. Here, we demonstrate, through experiments and simulations, how point-spread function engineering can be used to extend the reach of differential dynamic microscopy. 
\begin{singlespace}
$\textcopyright$ 2017 Optical Society of America. One print or electronic copy may be made for personal use only. Systematic reproduction and distribution, duplication of any material in this paper for a fee or for commercial purposes, or modifications of the content of this paper are prohibited.
\end{singlespace}
\end{abstract}

\vspace{10mm}
%\setboolean{displaycopyright}{true}

How molecules or small tracer particles move through biological or other soft materials, whether diffusively, ballistically or otherwise, is an important characteristic of these systems. Methods to characterize such dynamics include single-particle tracking \cite{mason_particle_1997}, utilizing real-space image data, and dynamic light scattering (DLS) \cite{berne_dynamic_2000}, using reciprocal space data. Combining elements of both real and reciprocal space methods is, among related digital Fourier microscopy techniques, differential dynamic microscopy (DDM) which, since its initial description in 2008 \cite{cerbino_differential_2008}, has been used to characterize the dynamics of bacteria \cite{wilson_differential_2011, martinez_differential_2012}, colloids \cite{giavazzi_simultaneous_2016}, DNA \cite{wulstein_light-sheet_2016} and protein clusters \cite{safari_differential_2015}. By analyzing in the Fourier domain a time series of real-space images one can use DDM to extract the temporal decay rate of density fluctuations within the sample across a range of wave vectors providing data analogous to that of DLS. A notable feature of DDM is that such real-space images can be acquired with a variety of microscopy techniques (e.g., dark-field \cite{bayles_dark-field_2016}, confocal \cite{lu_characterizing_2012}, light-sheet \cite{wulstein_light-sheet_2016}) making this method widely accessible. Here, we show that DDM can be extended with point-spread function (PSF) engineering. 

It has long been understood that the PSF of an imaging system can be advantageously engineered by modifying the electromagnetic field distribution in the system's pupil or Fourier plane with an appropriate filter. Applications of PSF engineering (PSFE) have grown with advances in microscopy such as single-molecule localization and two-photon microscopy and along with advances in and the increased availability of adaptive optical elements such as deformable mirrors \cite{perreault_adaptive_2002} and spatial-light modulators \cite{maurer_what_2011}. PSFE has been used to improve the resolution in confocal microscopy \cite{hegedus_superresolving_1986}, extend a microscope's depth of field \cite{dowski_extended_1995}, and enhance the point localization precision in all three dimensions \cite{quirin_optimal_2012}. As we show here, PSFE, already established as a useful tool for extracting information from real-space imaging, can be useful in digital Fourier microscopy methods like DDM. 

With DDM one is forced to make trade-offs, which are by no means unique to DDM, between maximizing the range of spatial frequencies or wave vectors ($q$) and the range of time scales covered. For instance, capturing fast dynamics requires high camera frame rates which often prohibits a large field-of-view. The range of spatial frequencies accessible is determined, on the high end, by the pixel size and, on the low end, by the largest dimension of the image. In practice the range is also influenced by the signal-to-noise ratio and the range of time scales probed. For instance, low-$q$ dynamics will, for diffusive motion, decay the slowest and therefore require extended data acquisition times. Conversely, high-$q$ dynamics require, for many instances, fast frame rates to capture. Finally, the signal-to-noise, drift, the microscope’s transfer function and other considerations can limit the spatial and temporal scales accessible.

Here, we demonstrate experimentally how PSFE can lessen the loss of low-$q$ dynamics when the imaged region-of-interest (ROI) is reduced. We fruitfully use astigmatism to prolong the duration in which a particle’s PSF is visible in the ROI when the ROI is small. We then demonstrate PSFE using simulations to highlight how dynamics in the axial dimension can be probed using DDM and an appropriate PSF. 

The principle and work-flow of DDM is explained and derived from fundamental imaging concepts in detail by others \cite{giavazzi_scattering_2009,giavazzi_digital_2014}. Briefly, to use DDM a time series of images are acquired. Images are then subtracted to generate a difference signal, $\Delta I(x,y;\Delta t) = I(x,y;t) - I(x,y; t+\Delta t)$.The Fourier power spectra of all differences corresponding to a given lag time, $\Delta t$, are computed and averaged to generate what is referred to as the image structure function, $D(\bm{q}, \Delta t)$, where $\bm{q} = (q_{x},q_{y})$ is the wave vector. For isotropic dynamics the radially averaged image structure function $D(q,\Delta t)$ is then fit to a function of the form
\begin{equation}
D(q,\Delta t) = A(q)[1-f(q,\Delta t)] +B(q).
\label{Eq1}
\end{equation}
The amplitude term, $A(q)$, depends on the structure of the sample and the microscope's optical transfer function. The background term, $B(q)$, depends on noise in the system. Sample dynamics are accounted for in the intermediate scattering function (ISF), $f(q, \Delta t)$. For normally diffusing and non-interacting particles the ISF takes the form, $f(q,\Delta t) = \exp{(-\Delta t/\tau(q))}$. The $q$-dependence of the decay time, $\tau(q)$, reveals the diffusion coefficient, $D$, of the particles with $\tau(q)=(Dq^{2})^{-1}$. 

\begin{figure}[!t]
\centering
\includegraphics[width = 0.75 \textwidth]{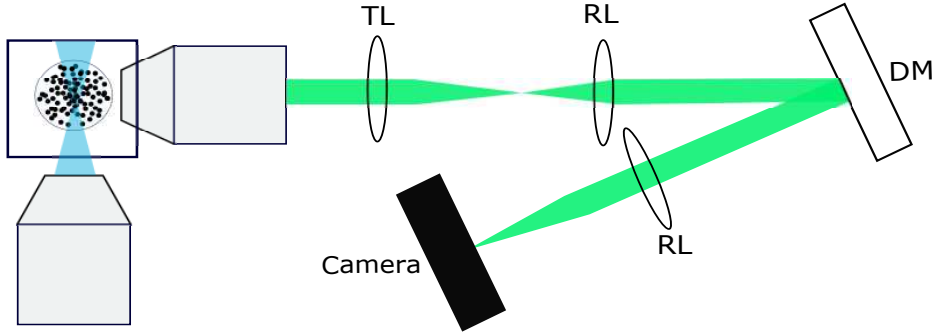}
\caption{Schematic of the imaging pathway. TL: tube lens; RL: relay lens; DM: deformable mirror.}
\label{Layout}
\end{figure}

Our experimental setup is shown in Fig. \ref{Layout} and details of the setup without the adaptive optics element can be found in Ref. 7. Briefly, we image samples of 200 nm fluorescent beads diluted in water (volume fraction $\approx10^{-6}$). Our imaging path consists of a 20$\times$ 0.5 NA water-dipping objective followed by a tube lens and an f=200 mm relay lens. Together these lenses image the back focal plane of the objective onto a piezoelectric deformable mirror (DMP40, Thorlabs) consisting of 40 actuators. The deformable mirror directs emitted light through the final relay lens (f=200 mm), and onto the camera (Zyla 4.2, Andor).

We first show how PSFE can be used to recover the range of wave vectors probed with DDM when reducing the imaging ROI. Decreasing the camera's ROI may be beneficial for DDM experiments as, for many cameras, it allows for higher frame rates. Unfortunately, a reduced ROI will also limit access to low-$q$/long-time dynamics. However, with PSFE we avoid sacrificing those dynamics when reducing the ROI for fast imaging.

While the loss of low-$q$ information upon reducing the observation volume has been noted before \cite{lu_characterizing_2012, wulstein_light-sheet_2016}, we demonstrate how reducing one dimension of an ROI limits the range of wave vectors one can probe. As the ROI is reduced, particles leave the observation volume quicker and, therefore, long-time dynamics are lost. We recorded 5000 images at 330 Hz with a resolution of 512$\times$128 pixels, corresponding to 128$\times$32 $\mu$m$^2$. We used DDM to analyze ROIs of 128$\times$128, 128$\times$32 and 128$\times$8 pixels and generate $D(q,\Delta t)$ as described above. In all cases, the full captured image series of 512$\times$128 was divided into as many non-overlapping ROIs as possible. We then averaged $D(q,\Delta t)$ from each subregion. In this way, the same total data is utilized regardless of the size of the subregion. 

\begin{figure}[htbp]
\centering
\includegraphics[width = 0.75 \textwidth]{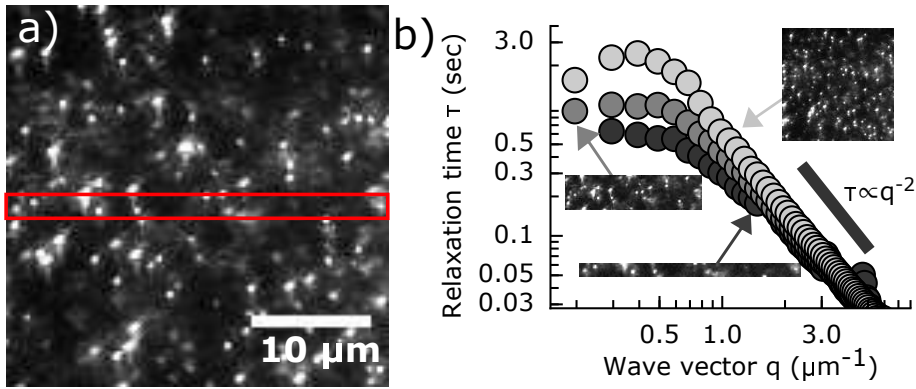}
\caption{(a) 128$\times$128 image. Rectangle shows the size of a 128$\times$8 ROI. (b) Relaxation times versus $q$ for ROIs: 128$\times$128 (lightest gray), 128$\times$32 and 128$\times$8 (darkest gray). We find a diffusion coefficient of $\approx$1.8 $\mu$m$^2$/s which would predict a hydrodynamic radius of $\sim$120 nm using the Stokes-Einstein relation.}
\label{Fig2}
\end{figure}

The observed trend of loss of low-$q$ dynamics with decreasing ROI is presented in Fig. \ref{Fig2}. On a plot of the relaxation time, $\tau$, versus $q$ a plateau of $\tau$ signifies that slower dynamics are inaccessible. We expect the low-$q$ plateau time to correspond to $\tau \simeq (d_{s}/2)^2/2D_{0}$, where $d_{s}$ corresponds to the smallest dimension of the imaged volume. For the 128$\times$8 ROI, $d_{s}=4\mu$m and we would expect a plateau time of $\sim$ 1 s, close to the observed plateau of $\sim$ 0.6 s. For the 128$\times$128 ROI, $d_{s}$ corresponds to the illumination light-sheet thickness of several microns, consistent with the measured plateau time of $\sim$ 2 s.

\begin{figure}[h!]
\centering
\includegraphics[width = 0.75 \textwidth]{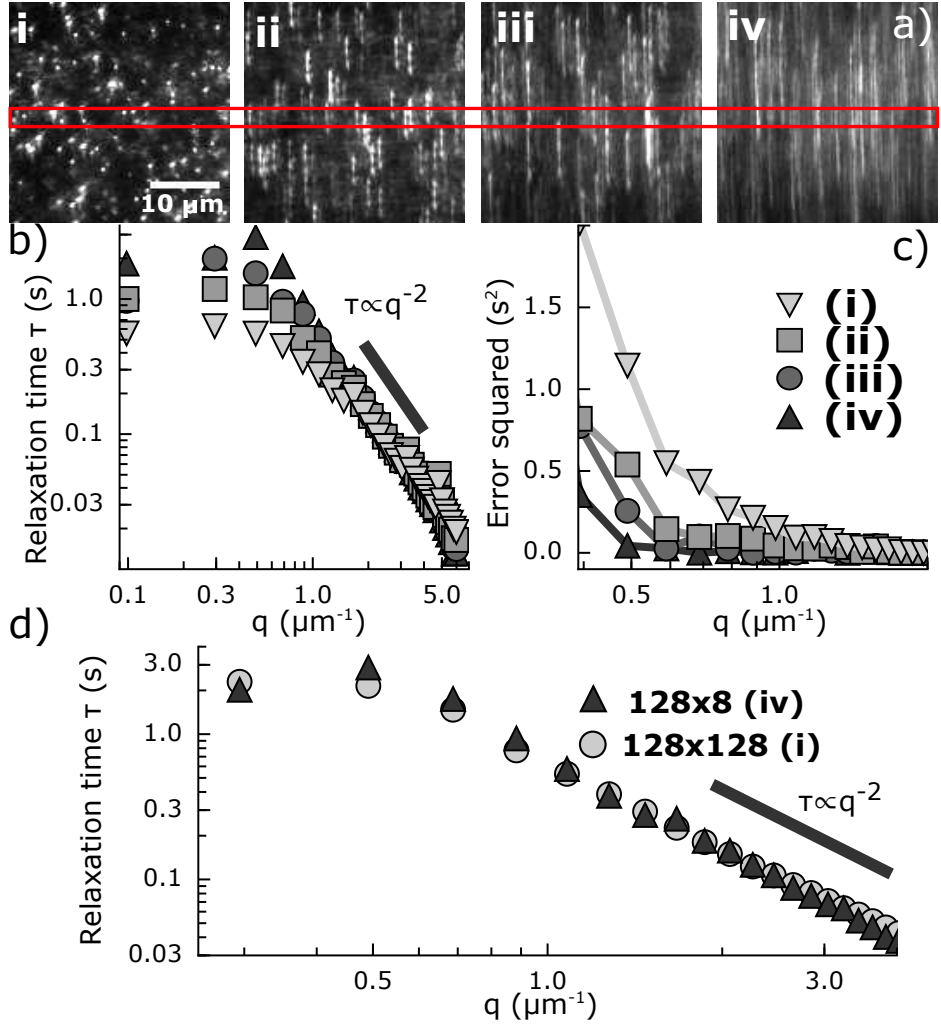}
\caption{ (a) 128$\times$128 ROI with increasing astigmatism from (i) to (iv). Rectangle demarks the 128$\times$8 ROIs. (b) Plot of $\tau$ versus $q$ for each level of astigmatism. (c) The deviation squared from the expected $\tau \propto q^{-2}$ line shows the low $q$ dynamics are better captured with increased astigmatism. (d) With the greatest astigmatism the accessible $q$ values using 128$\times$8 matches that of the full 128$\times$128 ROI with a standard PSF. }
\label{Fig3}
\end{figure}

To mitigate the effects of particles diffusing outside the imaging region we introduce an astigmatic aberration using the deformable mirror. When our astigmatic PSF is focused along the $x$-axis, the PSF along $y$ is extended up to several microns depending on the degree of astigmatism, as shown in Fig.\ref{Fig3}(a). With this purposefully introduced aberration we expected to recover the low-$q$ dynamics; a particle diffusing out of the size of the ROI, with a PSF elongated along the short dimension of the ROI, will not necessarily mean that its image has left the field of view. For instance a particle may quickly diffusive out of the 4 $\mu$m short dimension of the ROI indicated in Fig. \ref{Fig2}(a) but its extended image may remain detectable much longer provided it does not diffusive out of the excitation light.

Three astigmatic PSFs and the standard PSF are used to characterize how low-$q$/long-time dynamics are recovered. Movies are captured and analyzed as described above, with a total of 64 non-overlapping 128$\times$8 regions per recorded image sequence. With the elongation of the PSFs, the plateau increases from 0.6 s to 2.9 s, Fig. \ref{Fig3}(b). The recovery effectiveness is seen in the direct comparison between the 128$\times$128 regions with the standard PSF and the 128$\times$8 regions with the greatest astigmatism, Fig.\ref{Fig3}(d).   

We now turn to how PSFE allows for more fully capturing three-dimensional (3D) dynamics. For most microscopy setups and situations where DDM may be employed, the decay of the ISF due to dynamics along the optical axis can be neglected. A thorough discussion of the weighting of in-plane to out-of-plane dynamics for DDM measurements can be found in \cite{giavazzi_scattering_2009, giavazzi_digital_2014}. But briefly, unless the depth-of-field is highly confined, the time scale associated with in-plane intensity fluctuations is distinct from and much smaller than the time-scale associated with out-of-plane fluctuations at a given $q$.

In the case of fluorescence imaging, this can be understood by considering the shape of the 3D PSF. As the PSF is typically extended in the axial direction and symmetric about the focal plane, intensity fluctuations in the image plane will more strongly reflect in-plane rather than out-of-plane displacements. As a thought experiment, we can imagine a different PSF where the intensity in the image plane translates in $xy$ (instead of merely blurring) when the point source translates in $z$. If such a PSF could be realized then the observed in-plane intensity fluctuations would relate to the axial dynamics of the sample. We could then use DDM to find the time scale of intensity fluctuations and knowledge of the PSF’s geometry to relate those intensity fluctuations to density fluctuations in the sample.

Efforts to advance particle-tracking and single-molecule localization techniques have confronted this issue of how to maximize the sensitivity of the observed $xy$ slice of the PSF to the point-source’s position in $z$. Of particular relevance here, we highlight the use of PSFE to create greater and more easily distinguishable changes in the image that result from axial displacement of single emitters. For example, PSFs that are astigmatic \cite{kao_tracking_1994}, exhibit a double-helix shape \cite{pavani_three_2008} or result from a saddle-shaped pupil function \cite{schechtman_optimal_2014} have all been used to enhance axial localization precision.  

\begin{figure}[htbp]
\centering
\includegraphics[width = 0.75 \textwidth]{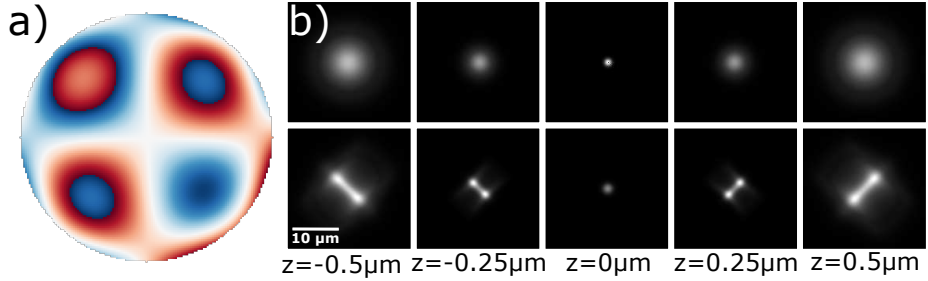}
\caption{(a) The pupil function of the SP-PSF used in the simulations. (b) Comparison between standard (top row) and SP (bottom row) PSFs.}
\label{Fig4}
\end{figure}

We show here that such concepts of PSFE can be used to enhance the sensitivity of DDM to 3D dynamics. We use what is referred to as a saddle-point PSF (SP-PSF) (Fig. \ref{Fig4}) \cite{schechtman_optimal_2014}. Such a PSF was designed to maximize the information that could be extracted about a point-like particle's position in the presence of high background. Whereas the standard PSF blurs and will become buried in noise upon defocus, the SP-PSF has two distinct lobes whose distance and angle vary with defocus. Thus, we speculated that the SP-PSF could help to extract dynamic information about particles moving in 3D with DDM.

We first simulated time series of images of point-like particles drifting at a uniform velocity in the axial direction. The field of view was about 21x21 $\mu$m$^2$ and approximately 40 particles were visible in each frame. We simulated 2000 images for each time series at a variety of particle drift speeds (9.6 nm/ time step to 38 nm / time step) and with both the standard and the SP-PSF. Those time series were analyzed using DDM as described previously and the image structure function was fit to: 

\begin{equation}
f(q, \Delta t) = \sin(Z\tan ^{-1}\theta)/[Z\theta (1+\theta^2)^{Z/2}],
\label{Eq2}
\end{equation}
where $\theta = \Delta t/(\tau(q)[Z+1])$. $Z$ characterizes the distribution of velocities and $\tau$ is the characteristic time related to the mean velocity via $\tau = 1/q \overline{v}$ \cite{germain_differential_2016}. 

Though this model for ballistic motion assumes velocities randomly directed in $xy$, we adopt it for our situation of ballistic motion entirely in $z$ since features of the SP-PSF in $xy$ change noticeably with motion in $z$. Additionally, we use a model that describes a distribution of velocities because the spreading of the SP-PSF in $x$ and $y$ with defocus is not constant with $z$-position.

\begin{figure}[h!]
\centering
\includegraphics[width = 0.75 \textwidth]{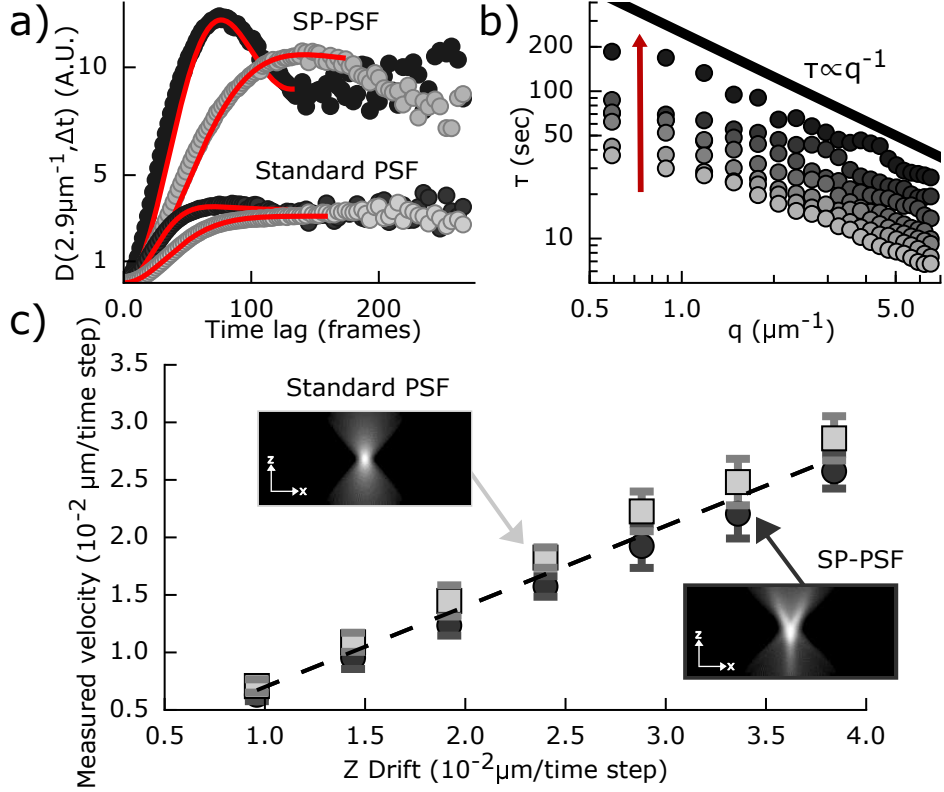}
\caption{(a) $D(q,\Delta t)$ for the standard and SP-PSFs plotted to compare the shapes and signal-to-noise all for $q$ = 2.9 $\mu$m$^{-1}$. The black symbols correspond to a simulated drift speed of 33.6nm/time-step and the lighter gray symbols to 19.2nm/time-step. (b) $\tau$ versus $q$ for the SP-PSF simulations. Arrow points towards increasing drift speed. (c) Measured velocity plotted versus the known $z$-drift for both PSFs. Inset images: $z$-$x$ slices of PSFs label corresponding data. }
\label{Fig5}
\end{figure}

Both the simulations using the standard and the SP-PSF fit to Eq. \ref{Eq2} and in both cases the measured mean velocity from that equation increases linearly with increasing simulated drift (Fig. \ref{Fig5}). As expected, the value of the measured velocity is not equal to the actual simulated drift velocity. Instead, the measured velocity is about 70\% of the actual simulated velocity. We found that while the ratio of the apparent to actual velocity is similar for both the standard and the SP-PSF, the signal-to-background ratio of the image structure function (measured by $A(q)/B(q)$) is greater by about a factor of 3 in the case of the SP-PSF (Fig. \ref{Fig5}(a)). Notable also is the oscillation of $D(q,\Delta t)$ identifiable when using the SP-PSF but not the standard PSF.

As with the case of particle localization, we expect the utility of the SP-PSF to emerge when images contain a high background. Therefore, we simulated particles again drifting along $z$ but also diffusing and in the presence of high background ($10^5$ signal photons per frame per particle with 3600 background photons per pixel). The simulated drift was 250 nm/time step in $z$ and a diffusion coefficient of 1150 nm$^2$/time step. 

\begin{figure}[htbp]
\centering
\includegraphics[width = 0.75 \textwidth]{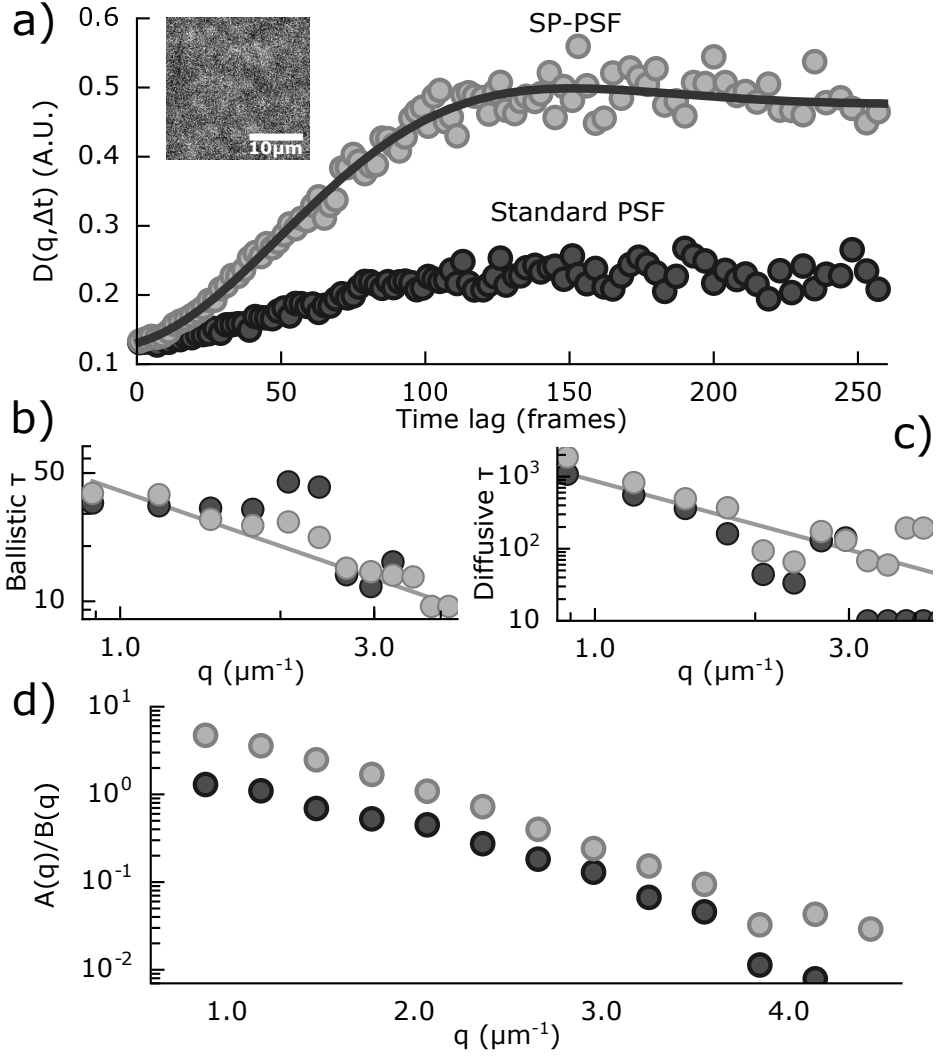}
\caption{For all plots, data from simulations using the SP-PSF are shown in light gray and the standard PSF in dark gray. Note that we used the same number of signal and background photons for simulations using each PSF. (a) $D(q,\Delta t)$ plotted versus $\Delta t$ for $q=1.8\mu m^{-1}$. Inset: a frame from simulation. (b,c) The characteristic ballistic time (b) and diffusive time (c) from fits to Eq. \ref{Eq1} using the $f(q,\Delta t)$ in Eq. \ref{Eq3}. The line shows the $\tau\propto q^{-1}$ scaling (b) and the $\tau=1/Dq^{2}$ relationship (c). (d) Signal to background determined by the ratio of the amplitude and background parameters of Eq. \ref{Eq1}. The number of photons emitted per particle per frame is equivalent for both simulations with the standard PSF and SP-PSF.}
\label{Fig6}
\end{figure}

As with the simulations without noise, we see that the signal to background ratio of $D(q,\Delta t)$ is better using the SP-PSF (Fig. \ref{Fig6}). With diffusion and drift present we fit the ISF to 

\begin{equation}
f(q, \Delta t) = \frac{\exp{(-\Delta t/\tau_d(q))} \sin(Z\tan ^{-1}\theta)}{[Z\theta (1+\theta^2)^{Z/2}]},
\label{Eq3}
\end{equation}
where $\tau_d$ is the characteristic diffusive time and the characteristic ballistic time is contained within $\theta$ as previously shown in Eq. \ref{Eq2}. From the fits of the simulated data using the SP-PSF we see that the characteristic ballistic time follows the expected $\tau \propto q^{-1}$ relationship to a much greater extent than the standard PSF simulations (Fig. \ref{Fig6}(b)). The characteristic diffusive time's expected scaling of $\tau_d \propto q^{-2}$ is also more evident with the SP-PSF as opposed to the standard PSF (Fig. \ref{Fig6}(c)). For the majority of wave vectors where we could fit $D(q,\Delta t)$, the signal to background ratio is greater by about a factor of 3 using the SP-PSF (Fig. \ref{Fig6}(d)).

A strong selling point of DDM is its accessibility. It allows one to acquire DLS-type data with a simple optical microscope (of a variety of modalities) without requiring expertise in advanced optical instrumentation. Conversely, PSFE with adaptive optics elements is likely beyond the expertise of many material science or biology labs. Fortunately, the PSFE discussed here could be accomplished with static optical elements in configurations simpler than ours. The astigmatic PSF could be realized with a simple cylindrical lens in the imaging path. Less straightforward would be implementing the SP-PSF, but a static phase mask could be fabricated. 

We have shown, through experiments and simulations, that PSFE can enhance the capabilities of the digital Fourier microscopy technique DDM. In particular, with PSFE we have enhanced DDM's sensitivity to dynamics in 3D and mitigated the detrimental effects of limiting the camera's ROI. We hope that continued developments in PSFE, spurred by advances in other microscopy techniques and in the growing knowledge base covering adaptive optics, in concert with advances in digital Fourier microscopy will provide researchers with a growing collection of methods to interrogate sample dynamics over expanding ranges of spatial and temporal scales.

\vspace{12mm}
National Institutes of Health (1R15GM123420). We acknowledge the Donors of the American Chemical Society Petroleum Research Fund for partial support under award 57326-UNI10. 

We thank Yoav Shechtman (Technion – Israel Institute of Technology) for providing the SP-PSF used in simulations and for helpful discussions. 

%\section{References}
% Bibliography
\iffalse

\fi
\bibliography{bibb.bib}

\end{document}